\documentclass[conference]{IEEEtran}

\usepackage{comment}
\usepackage{float}
\usepackage{siunitx}

\ifCLASSINFOpdf
   \usepackage[pdftex]{graphicx}
\else
\fi
\usepackage{amsmath}
\usepackage{array}

\usepackage{listings}
\usepackage{caption}
\usepackage{subcaption}
\usepackage{svg}
\usepackage[T1]{fontenc}
\usepackage{inconsolata}

\usepackage[sorting=none]{biblatex} % Utiliza biblatex y ordena por orden de aparición
\usepackage{comment}
\usepackage{color,soul}

\begin{document}
\title{Implementing Multi-GPU Scientific Computing Miniapps Across Performance Portable Frameworks}

\author{
\IEEEauthorblockN{Johansell Villalobos}
\IEEEauthorblockA
{
    National High Technology Center \\
    San José, Costa Rica\\
    jovillalobos@cenat.ac.cr \\
    0009-0002-3398-0714
}
\and
\IEEEauthorblockN{Josef Ruzicka}
\IEEEauthorblockA
{
    National High Technology Center and \\
    Costa Rica Institute of Technology  \\
    San José, Costa Rica \\
    jruzicka@cenat.ac.cr \\
    0009-0003-4423-2612
}
\and
\IEEEauthorblockN{Silvio Rizzi}
\IEEEauthorblockA{
    Argonne National Laboratory\\ 
    Lemont, Illinois \\
    srizzi@anl.gov \\
    0000-0002-3804-2471
}
}


%Telephone: (800) 555--1212, Fax: (888) 555--1212}


\maketitle

\begin{abstract}
    Scientific computing in the exascale era demands increased computational power to solve complex problems across various domains. With the rise of heterogeneous computing architectures the need for vendor-agnostic, performance portability frameworks has been highlighted. Libraries like Kokkos have become essential for enabling high-performance computing applications to execute efficiently across different hardware platforms with minimal code changes. In this direction, this paper presents preliminary time-to-solution results for two representative scientific computing applications: an N-body simulation and a structured grid simulation. Both applications used a distributed memory approach and hardware acceleration through four performance portability frameworks: Kokkos, OpenMP, RAJA, and OCCA. Experiments conducted on a single node of the Polaris supercomputer using four NVIDIA A100 GPUs revealed significant performance variability among frameworks. OCCA demonstrated faster execution times for small-scale validation problems, likely due to JIT compilation, however its lack of optimized reduction algorithms may limit scalability for larger simulations while using its out of the box API. OpenMP performed poorly in the structured grid simulation most likely due to inefficiencies in inter-node data synchronization and communication. These findings highlight the need for further optimization to maximize each framework's capabilities. Future work will focus on enhancing reduction algorithms, data communication, memory management, as wells as performing scalability studies, and a comprehensive statistical analysis to evaluate and compare framework performance.

\end{abstract}


% \fi
\IEEEpeerreviewmaketitle

\section{Introduction}\label{sec:intro}
% Scientific computing workflows
% Multi-GPU accelerated workflows
% state of the art of HPC systems
% differences between HPC systems
% Performance portability

The modeling and simulation of complex physical phenomena remain central to scientific computing, particularly with the aid of high-performance computing (HPC) systems. As we advance into the exascale era, increasingly refined and complex models of reality are becoming feasible, which in turn demand more sophisticated computational methods and efficient utilization of HPC resources. These simulations, often involving intricate numerical methods for particle interactions or field variable evolution, require substantial computational power and advanced algorithms to effectively leverage heterogeneous parallel architectures.

Simulations typically rely on specific spatial and temporal scales to ensure numerical stability and convergence. These constraints can limit the ability to conduct large-scale simulations, essential for comprehensive analysis of the system being modeled. HPC systems address this challenge by enabling the distribution of data across multiple compute nodes, thereby enhancing simulation efficiency and allowing for more detailed investigations.

The conventional approach for workload distribution in HPC is the Message Passing Paradigm, which leverages the memory and compute power of multiple nodes. Each node may also employ shared memory for intra-node parallelism, with some systems optimized for multi-core CPUs and others for GPU accelerators. The latter, particularly with general-purpose GPUs (GPGPUs), has become increasingly prevalent due to their ability to perform parallel floating-point arithmetic efficiently. However, programming for GPUs introduces complexities, such as the need for specific memory access patterns and architecture-specific libraries like CUDA, HIP, and SYCL.

To address these challenges, performance-portable frameworks have been developed, allowing a single source code to run efficiently across different architectures, regardless of the hardware's type or vendor. This paper investigates four of the leading frameworks in performance-portability—OpenMP, Kokkos, RAJA, and OCCA—by implementing two representative scientific miniapps: an N-body system and a structured grid simulation. We use the Message Passing Interface (MPI) for inter-node scalability on nodes equipped with four NVIDIA A100 GPUs. This work provides insights into the development of performance-portable codes, along with runtime evaluations. The remainder of the paper is organized as follows: Section~\ref{sec:miniapps} discusses the miniapps; Section~\ref{sec:perfport} refers to the performance portability frameworks; Section~\ref{sec:relwork} elaborates on related work; Section~\ref{sec:impl} details our implementation; Section~\ref{sec:res} presents our preliminary results; and Section~\ref{sec:final} offers conclusions and future work directions

\section{Miniapps}\label{sec:miniapps}

\subsection{N-Body Simulation}
    The N-Body simulation is a computational method that is implemented for the study of the dynamics of systems of particles or bodies under physical forces such as gravity or interatomic potentials. Widely applied in fields like astrophysics, computational biology, and computational chemistry, N-body simulations are crucial for examining both large-scale phenomena, such as galaxy dynamics, and smaller-scale systems in materials science \cite{LAMMPS,Springer2012BerkeleysDO}. These simulations provide key insights of phenomena that otherwise would be impossible to capture, hence the importance of this method and our choice of application.
    
    %N-body simulations are often implemented in the fields of astrophysics, computational biology, and computational chemistry, in applications that study the evolution of large scale systems such as the dynamics of galaxies to smaller scale systems as seen in materials science \cite{LAMMPS,Springer2012BerkeleysDO}. 

    In our case we implemented a Lennard-Jones gas as our N-Body system. The Lennard-Jones gas is governed by the following interatomic potential and force,
    \begin{align}
        \begin{split}
        V(r_{ij}) &= 4 \epsilon \left[ \left( \frac{\sigma}{r_{ij}} \right)^{12} - \left( \frac{\sigma}{r_{ij}} \right)^{6} \right] \\
        \textbf{F} = -\nabla V &= 24 \epsilon \left[ 2 \left( \frac{\sigma^{12}}{r_{ij}^{13}}\right)-\left( \frac{\sigma^{6}}{r_{ij}^{7}}\right)\right] \hat{\textbf{r}}.
        \end{split}
        \label{eq:ljpotential}
    \end{align}

    \noindent where $r_{ij}$ is the distance from particle $i$ to particle $j$. The force is then calculated for each pair of particles in the system and the position of each particle is updated using the Verlet half-stepping algorithm.  

\subsection{Structured Grid Simulation}
    Structured grid simulations are computational methods that usually solve partial differential equations (PDEs), with the use of discretizations such as the finite difference, volume or element method in a regularly spaced, predictable data layout. These simulations often lead to better memory usage due to the predictability of memory access patterns \cite{Springer2012BerkeleysDO}. These simulations can also be parallelized more efficiently in HPC systems. Some downsides of structured grid simulations are that they are geometrically stiff, as opposed to unstructured grid methods. Structured grid methods are used to solve problems in a variety of fields ranging from solid mechanics to climate modeling. 

    For our structured grid simulation we implemented a two-dimensional fluid dynamics simulation using the vorticity formulation of the Navier-Stokes equation, 
    \vspace{-0.2in}
    
    \begin{align}
        \begin{split} 
            u = \frac{\partial \psi}{\partial y} \quad v = &-\frac{\partial \psi}{\partial x}; \quad\quad
            \nabla^2 \psi = -\omega\\
            \frac{\partial \omega}{\partial t} + &u\frac{\partial\omega}{\partial x}+v\frac{\partial \omega}{\partial y} =\nu\nabla^2 \omega 
        \end{split}
        \label{eq:vorticity}
    \end{align}
    
    \noindent in which $\psi$ is the fluid stream function, $u$, $v$ represent both components of the velocity of the fluid, $\omega = \nabla \times \textbf{u} = (0,0,\omega)$ is the vorticity of the fluid and $\nu$ is the kinematic viscosity of the fluid. This equation is evolved using forward Euler time integration and a central difference spatial discretisation in a uniform Cartesian grid.

\section{Performance Portability Frameworks}\label{sec:perfport}
%Ensuring that applications run efficiently on different hardware platforms is crucial for advancing research in fields requiring large-scale computations. Multiple frameworks have been, and are currently being developed to tackle the limitations in portability presented by older, architecture-specific frameworks. 
%In this work we make use of four state-of-the-art performance portability frameworks: Kokkos, RAJA, OCCA, and OpenMP, all of which are detailed in this section. 

Efficient execution of applications across diverse hardware platforms is vital for advancing research in fields reliant on large-scale computations. To address the portability limitations of traditional frameworks, performance-portable frameworks have evolved, with ongoing development enhancing their capabilities. This work leverages four state-of-the-art performance portability frameworks—Kokkos, RAJA, OCCA, and OpenMP—each detailed in this section, to ensure optimal performance across architectures.

\subsection{Kokkos}
    Currently under active research and development, the Kokkos C++ library is designed to address architectural differences in memory access patterns. This is achieved through the use of \textit{Kokkos Views}, which are multidimensional arrays with \textit{polymorphic data layout} which reside in a \textit{memory space}. This data storage abstraction tackles the \textit{array of structures} (AoS) versus \textit{structure of arrays} (SoA) problem, which refers to the fact that a parallel computational kernel requires to follow a blocked access pattern when its execution lies within a CPU, and a coalesced data access pattern in order to optimize its performance.
    
    Additionally, Kokkos provides abstractions for fine-grain data parallelism across manycore architectures in a library approach \textit{i.e.}, without the use of directives or source-to-source translators. In this abstraction, threads execute their respective kernels in an \textit{execution space}. To unify its memory access and fine-grain data parallelism abstractions, Kokkos maps its code to the optimal back-end for each target device using thin back-end implementations~\cite{kokkos}.
    
\subsection{RAJA}

    RAJA, like Kokkos, provides a portability and abstraction layer that allows a single source code base to work across various architectures and programming models. It supports CUDA, HIP, SYCL, and OpenMP backends on GPUs and aims to maintain non-invasive code by being agnostic to memory management. RAJA also offers optional automatic memory management through RAJA's device abstraction layer  \texttt{RAJA::resources} interface to CAMP, although this adds complexity to the code base.
    
    This framework includes \texttt{RAJA::View} for managing data layout and access, which simplifies indexing and supports layout permutations, offsets, and strided indexing, with bounds checking for debugging \cite{rajaRAJA}. In simpler implementations, like the ones in this work, data layout was managed directly within parallel execution loops rather than using \texttt{RAJA::View}.
    
    RAJA's execution model relies on policies that define how and where loop kernels execute. Policies range from simple directives to complex nested ones, facilitating parallelism across different architectures. RAJA provides loop constructs such as \texttt{RAJA::forall}, \texttt{RAJA::kernel}, and \texttt{RAJA::scan}, and uses reduction objects (\texttt{RAJA::ReduceSum}, \texttt{RAJA::ReduceMax}, etc.) for performing reductions, which can be used within these constructs for flexible computation.

%RAJA similar to Kokkos is a portability and abstraction layer that enables the exploitation of various architectures and programming models using a single source code base \cite{Beckingsale2019}. RAJA focuses on providing performance portability with the CUDA, HIP, SYCL and OpenMP backends on GPU. This portable code abstraction layer was developed with memory management agnosticism, which is in the direction of the specific goal of maintaining the non-invasivity of code. However, RAJA also provides automatic memory management by interfacing with the CAMP library, specifically \texttt{RAJA::resources} abstraction for devices
    
\subsection{OCCA}
    OCCA (Open Concurrent Computing Abstraction) is a framework aimed at heterogeneous programming on GPUs and field-programmable arrays (FPGAs). This framework is vendor neutral as it supports the CUDA, HIP, SYCL, OpenMP, OpenCL and Metal backends \cite{githubOccadocsMain}. OCCA has the \texttt{occa::device} as well as the \texttt{occa::memory} class which abstract the memory managing process. An interesting aspect of this framework is that it has the capability of choosing a platform target at run-time due to just-in-time (JIT) code generation, which translates to being able to use different devices concurrently. OCCA implements a macro-based approach which is called OCCA kernel language (OKL). This approach enables the user to write specific kernels that are to be executed on the different backends available at runtime. Because of this we decided to manage memory using the \texttt{occa::device} class. Pointers to device memory are allocated using the \texttt{occa::device::malloc} function. 

    With respect to data layout, OCCA does not present a specific object or class as Kokkos or RAJA with their \texttt{View} objects. In-kernel data is treated similarly to data pointers in C/C++ loops.
    
    The execution model differs from the rest of the frameworks as backends and kernel definitions are set using JSON files at runtime. Parallel loops in OCCA are characterized by an added statement to a typical for-loop, which controls different levels of parallelism. Reductions can be performed by using \texttt{occa::memory} and the \texttt{@atomic} clause inside kernels. Similar to RAJA, these objects are independent of data traversal and enable the user to perform any number of reductions and computations in the kernel.
    
\subsection{OpenMP}

    The OpenMP (Open Multi-Processing) API specification for parallel programming~\cite{OpenMP2024} provides library routines and compiler directives in the form of \texttt{\#pragma} clauses, that allow high-level parallelism in C, C++, and Fortran applications designed for shared-memory multicore architectures. OpenMP remains one of the standard frameworks for performance portability due to its ease of use, which allows developers to parallelize code with minimal changes, with simple directives that address functionalities like GPU computation offloading and data transfer management, without the need for specifying execution or memory spaces. Its maturity, large community base, and frequent updates make it a reliable choice for developing parallel applications across various platforms. These characteristics have helped OpenMP maintain its relevance and popularity in the field of parallel computing.
    
    In contrast with how other frameworks handle data structures, OpenMP maps data to and from the devices using the same data handle, or "variable" representation of the memory, that way, the memory would be accessed depending on the device currently in use.

% Specific Miniapps
% Specific Frameworks

\section{Related work} \label{sec:relwork}

    When developing large code bases or porting legacy HPC codes, researchers often create a single code variant, limiting their ability to explore various programming frameworks across different hardware platforms. Consequently, studies focusing on problem abstractions and representations common to diverse research domains are essential for effective analysis and decision-making.
    
    Prior studies have highlighted valuable insights and benchmarks in performance portability. Martin \textit{et al.} \cite{martin} evaluated porting a multi-GPU fluid dynamics simulation from CUDA to SYCL, HIP, and Kokkos, finding that native models often perform better and are easier to port using automatic tools like HIPify \cite{hipify} and Intel’s DPCT \cite{dpct}. Kokkos, while the most portable, required manual porting and had performance trade-offs on some devices.
    
    Other research has further examined performance portability across various frameworks and architectures. Deakin \textit{et al.} \cite{deakin} conducted a broad study, showing that Kokkos and OpenMP provided the best portability for applications from the dwarves’ domain. Similar findings were reported in studies benchmarking Kokkos and SYCL on matrix-vector multiplications \cite{milc} and exploring OpenMP, OpenACC, CUDA, SYCL, and Kokkos on computational kernels for multi-material simulations \cite{materials}, with Kokkos and OpenMP again proving the most portable.
    
    Given this context, OpenMP and Kokkos stand out as critical choices for performance portability due to their widespread success across different studies. This paper extends the exploration of performance portability by incorporating RAJA and OCCA, offering additional insights into their potential within the field of scientific computing.

\section{Implementation}\label{sec:impl}
\subsection{General remarks}

    Our implementation of the simulations focused on leveraging the power of a single HPC compute node, by using a distributed memory approach along with GPUs to accelerate calculations. For this we used MPI in combination with the four performance portability libraries Kokkos, RAJA, OCCA, and OpenMP. We collected experimental data using the Polaris supercomputer at the Argonne Leadership Computing Facility (ALCF).

\begin{comment}
    \begin{table}
        \centering
        \caption{\hl{Polaris compute node configuration}}
        \begin{tabular}{c|c}
        \hline
            \textit{\textbf{CPU}}& 1 $\times$ (AMD Zen 3 Milan 2.8 GHz 7543P)  \\
            Cores & 64 \\
            RAM & 512 GB (DDR4) \\
            Storage & 3.2 TB (SSD) \\
            \hline
            \textit{\textbf{GPU}} & 4 $\times$ (NVIDIA A100 HGX) NVLink 600 GB/s \\
            GPU Memory & 4 $\times$ (40 GiB HBM2) \\
            GPU Memory BW & 1.6 TB/s \\ 
            Interconnect & NVLink 600 GB/s \\         
        \hline
        \end{tabular}
        \label{tab:polaris}
    \end{table}
\end{comment}
    \begin{figure}
        \centering
        \includegraphics[width=0.45\linewidth]{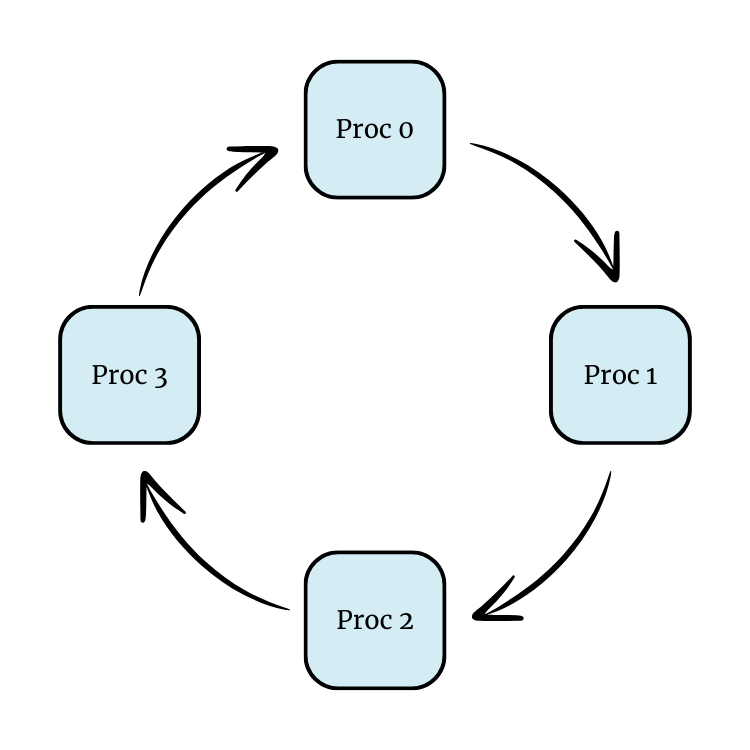}
        \caption{\centering MPI Ring exchange communication pattern implemented for the N-body simulation.}
        \label{fig:ringexchange}
    \end{figure}
    In our implementation we used GPU-aware MPI along the CUDA unified memory approach with the frameworks that supported it, being those RAJA, Kokkos, and OpenMP. We considered that this approach was the most likely to be implemented as high productivity is greatly valued in scientific computing \cite{Vetter2018}. As aforementioned, we implemented our own memory manager for the RAJA portability suite as it was the most straight forward way of allocating memory, and the common denominator found in the provided examples.

    To ensure the simulations are modeling the physics of the problem correctly we output the results using VTK (Visualization ToolKit) files that can be easily visualized using ParaView. The VTK outputs are per MPI rank meaning there will be $N_{ranks}$  files written to disk for visualization. As our aim is to execute the simulations for framework testing we consider that no visualization is needed after code and numerical method debugging.
    \vspace{-0.1in}
    \begin{figure}[H]
        \centering
        \includegraphics[width=0.5\linewidth]{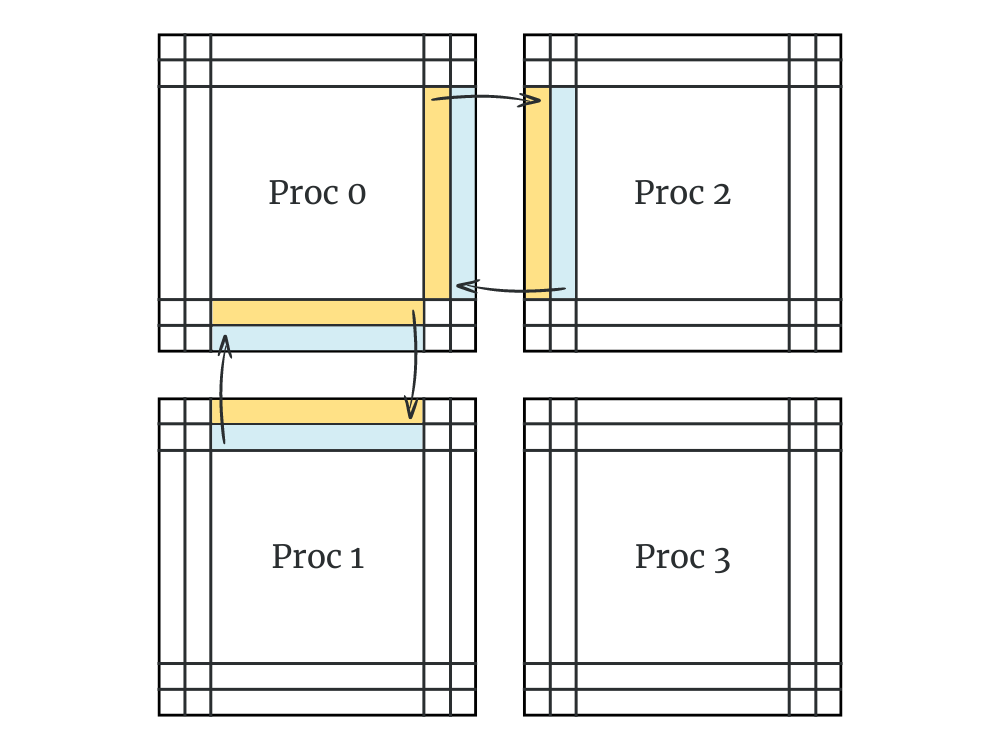}
        \caption{\centering MPI halo exchange communication pattern implemented for the structure grid simulation.}
        \label{fig:haloexchange}
    \end{figure}

    During the testing phase we implemented kernel timing using the \texttt{std::chrono} library to maintain uniformity in timing. Kernel timing was performed after host-device synchronization to ensure that the measurement accurately reflects the execution time of the kernel on the GPU without including any overhead from asynchronous operations or data transfers. It is important to consider that host-device synchronization is different for every framework. For Kokkos the \texttt{Kokkos::fence()} function is used, for OCCA the \texttt{occa::device::finish()} function is used, for RAJA we implemented our own \texttt{synchronize()} function which is dependent on the GPU architecture being used, in the present work this function calls  \texttt{cudaDeviceSynchronize()}. OpenMP, on the other hand, performs synchronization automatically after exiting a \texttt{target} scope so for this specific framework the we did not implement an explicit host to device synchronization function.

    Even if the frameworks abstract the concepts necessary for GPU execution, a degree of knowledge about GPU architecture is still necessary to implement blocked kernels in some of the frameworks. The concept of blocks and threads within a single stream processor is implemented both in RAJA and OCCA to ensure a better use of the architecture. Tiling is automated in both Kokkos and OpenMP target as they ensure optimization at compile time for the code with specific architectures, therefore choosing the best parameters for execution of code \cite{kokkos,Openmptargetpap}. %However, using the \texttt{#pragma omp target teams distribute parallel for} directive produced better results for our simulations than the more general \texttt{#pragma omp target teams loop}.

 \begin{comment}
    \begin{lstlisting}[
        language=C++, 
        label=lst:timing, 
        caption={\centering \hl{Device kernel timing, synchronization of device and host is performed differently across frameworks.}}, 
        captionpos=b]
...
    start = std::chrono::high_resolution_clock::now();
    device_func();
    synchronize_host_device();
    end = std::chrono::high_resolution_clock::now();
    duration += end - start;
...
    \end{lstlisting}
\end{comment}

\subsection{N-Body Simulation}
    
    Our implementation of N-Body method is characterized by performing force calculations between all particles which translates to $N_{p}^2$  operations where $N_p$ is the number of particles. Particle data was encoded as a contiguous array. Mass, position, velocity and force were stored, which resulted in a $N_p\times10$ array of doubles per processor. To take advantage of computational resources we parallelized our N-Body simulation by dividing the total amount of particles equally across all MPI ranks this way we ensured good load balancing during the simulation. We then implemented a ring exchange communication pattern to account for every particle present on each GPU. We used a periodic 1-dimensional Cartesian topology to implement the ring exchange communication. Figure \ref{fig:ringexchange} shows a diagram of the communication topology, in which we perform $N_{ranks}-1$ exchanges to obtain the complete force for each particle.

\subsection{Structured Grid Simulation}

    For the structured grid simulation we followed convention and stored our mesh points as one dimensional contiguous arrays in memory. For this simulation we stored arrays for $\psi$, $\omega$, $u$, and $v$. We implemented a halo exchange communication pattern by using a non-periodic 2-dimensional Cartesian topology. Figure \ref{fig:haloexchange} shows the diagram of the communication topology used. To solve for $\psi$ in the Poisson equation specified in equation \ref{eq:vorticity} we used Jacobi iteration and an explicit forward Euler advancing scheme to update $\omega$ through time.

% MPI parallelization
% RAJA, own memoryManager didn't use CAMP/UMPIRE devices
% Virtual memory
% VTK visualization for output checking
% Kernel Timing, fences, std::chrono why?,
% Talk about omp target teams parallel for
% Tiling
\section{Results}\label{sec:res}

% software stack
% Timing results
% 

    \begin{figure}
        \centering
        \includegraphics[width=0.75\linewidth]{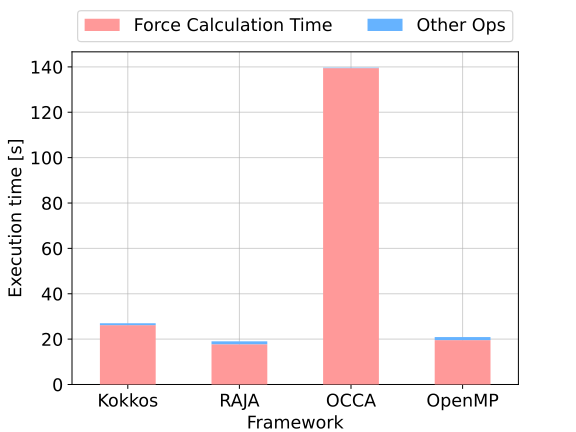}
        \caption{\centering Execution time results for the N-Body simulation across frameworks including reductions for energy calculations. Verlet integration, boundary checking and I/O are included as \textit{Other Ops}.}
        \label{fig:nbodyexecutionred}   
    \end{figure}
    
    Kernel timing and execution was conducted on a single Polaris node with 4 MPI ranks to leverage the node's 4 NVIDIA GPUs. In this work we considered small problem sizes, which served as validation for the correctness of the numerical methods, effectiveness of the GPU resource utilization for each framework, and effectiveness of memory distribution and communication before scaling up to larger problem sizes. For each simulation we performed a set of 10 repetitions to ensure we captured execution time variability. With respect to timing results, only the simulation loop section execution time was measured, omitting initialization. 
    
    The initial experimental kernel timing results for the N-Body simulation were collected using 10000 particles and 2500 iterations. Figure \ref{fig:nbodyexecutionred} shows a plot for the median execution time results for the N-Body simulation. These results exhibited the best time-to-solution obtained by RAJA with a median time of \SI{19.02}{\second}, closely followed-up by OpenMP's \SI{21.01}{\second}. Kokkos obtained \SI{26.96}{\second}, about 1.4 times slower than  RAJA's execution time. However the OCCA code variant presented a median time of \SI{139.82}{\second}, close to 7.3 times slower in relation to RAJA. 
    
\begin{figure}
        \centering
        \includegraphics[width=0.75\linewidth]{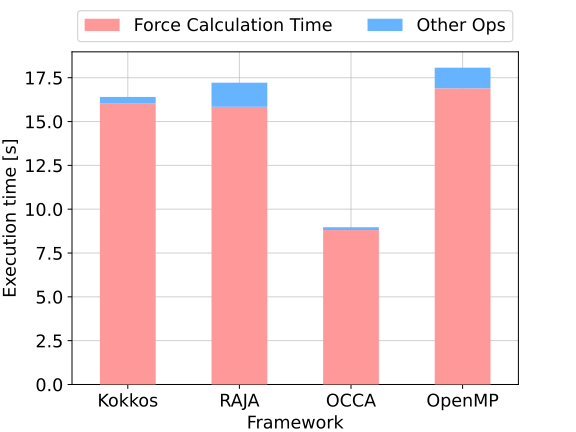}
        \caption{\centering Execution time results for the N-Body simulation across frameworks without reductions for energy calculations. Verlet integration, boundary checking and I/O are included as \textit{Other Ops}.}
        \label{fig:nbodyexecution}    
    \end{figure}

    This implementation of the N-Body simulation calculated the system's total energy by performing a reduction over all particles, employing equation \ref{eq:ljpotential}, within the force calculation kernel. This evidently would incur in an overhead in computation time depending on the reduction algorithm implemented in each framework. To inspect further into the non-negligible underperformance of the OCCA framework, we performed tests without the reduction operations on all of the frameworks, but rather as normal parallel loop constructs. This time, OCCA displayed results that surpassed the previous best observation achieved by RAJA, with \SI{8.96}{\second} median execution time, as observed in Figure~\ref{fig:nbodyexecution}. A possible reason for OCCA's improved performance may be that it is JIT compiled. This implies that hardware and program execution information could be used by the compiler to perform code optimizations based on hardware information that is not available for ahead-of-time compilers \cite{jitcomp}.    
    
    The OCCA library presents a considerable overhead which can be explained by the \texttt{@atomic} reduction used in the code. This is the only provided way to implement a reduction using OCCA's API and OKL. The alternative is to manually implement an optimized reduction algorithm within the force calculation kernel. From our perspective this is one of the drawbacks that the OCCA framework presents considering that all the other frameworks present already implemented and optimized versions of reduction algorithms.
    
\begin{figure}
        \centering
        \includegraphics[width=0.75\linewidth]{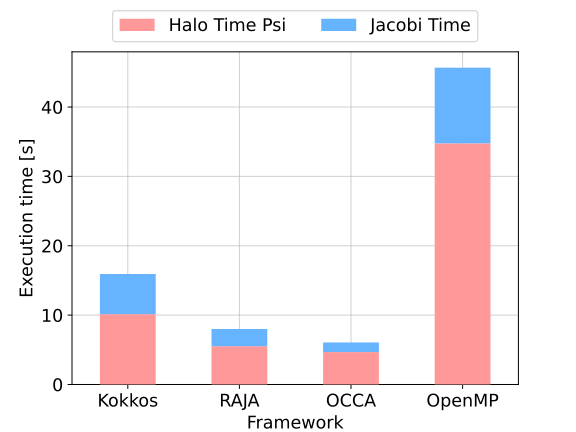}
        \caption{\centering Execution time results for the vorticity simulation across frameworks. Euler integration, boundary checking, and I/O times are negligible with respect to the halo exchange and Jacobi kernel times during the solution of the Poisson equation.}
        \label{fig:vorticityexecution}  
    \end{figure}

    The Structured Grid Simulation was tested on a $100\times100$ grid over 10 timesteps, with a tolerance of 0.0001 for the Jacobi iteration algorithm to gather the preliminary results showcased in this study. Although the Jacobi iteration algorithm is the computationally intensive part of our Structured Grid simulation, it has significant overhead in communication, represented as Halo Time Psi in Figure \ref{fig:vorticityexecution}. 
    
    In contrast with the tendency observed with the first N-Body simulation results, the OCCA code variant achieved the best time-to-solution performance with a median execution time of \SI{6.07}{\second}, surpassing RAJA's \SI{8.024}{\second}. In this case Kokkos finished third, with a duration of \SI{15.98}{\second}. Surprisingly, OpenMP obtained the poorest performance results, falling short of OCCA's execution time by approximately 7.5 times, with a duration of \SI{45.77}{\second}.
    
    These preliminary results suggest that further code optimizations should be written for our implementation variants in order for them to achieve comparable performance with the best-performing frameworks, as well as to understand the reason why OCCA and OpenMP displayed such underperformances, and why the Kokkos variant fell short of matching the best observed results. Additional to optimzing the data communication patterns and conducting performance analysis studies, it will be crucial to perform scalability experiments varying problem sizes, as the validation configurations presented in this study might prove to be too small for our chosen frameworks to fully exploit the hardware's potential with respect to parallelisation and data management capabilities.

% OCCA performs interestingly better
% Mention reductions @atomicsge
% 

% N = 10 reps, bash
% Report results with and without reduction

% Talk about OCCA and its
\section{Final Remarks}\label{sec:final}
In this paper, we have presented code implementations of two representative scientific computing applications: an N-body simulation and a structured grid simulation. These simulations were designed as miniapps for benchmarking purposes, implemented in C++ using a distributed memory approach, and hardware acceleration via four state-of-the-art performance portability frameworks: Kokkos, RAJA, OCCA, and OpenMP.

Our experimental results revealed variability in performance across different frameworks. Notably, OCCA, despite lacking optimized reduction algorithms, presented faster execution times for small scale problems compared to the other frameworks. This advantage in performance could be attributed to OCCA's JIT compilation, which may allow for more effective and informed optimization at runtime. However, the absence of optimized reduction operations in OCCA remains a limitation that should be addressed to improve its suitability for larger-scale, reduction-intensive simulations. In contrast, the OpenMP implementation showed poor performance for the structured grid simulation, primarily due to overhead in MPI communication, highlighting potential inefficiencies in handling inter-node communication and data synchronization.

%The experimental results indicate that there is variability within all performance portability frameworks. An interesting finding is that while reductions are not optimized in OCCA, it presents faster execution times than the other performance portable libraries at small scale problems which could be due to its JIT compiled nature. However we do consider optimized reduction algorithms a must in a performance portability framework. Our preliminary results also showed poor performance of OpenMP implementation of the structured grid simulation. We specifically note the overhead in MPI communication this code has which highlights potential inefficiencies in its handling of inter-node communication and data synchronization. 

While our preliminary results offer valuable insights into the behavior of the four frameworks, further optimization of the code implementations is crucial to fully exploit their capabilities. Future work should focus on optimizing reduction algorithms, enhancing data communication strategies, and fine-tuning memory management to improve overall performance. Additionally, scalability studies with larger problem sizes and different hardware platforms will be crucial to assess how well each framework scales and adapts to varying computational conditions. Furthermore, a profound statistical evaluation is needed to correctly compare performance portability results across frameworks.

%\hl{TODO: hablar de como estos resultados sirven para quién tenga que tomar una desición de diseño y no pueda analizar a priori cuál modelo le sirve más para su app específica, esto lo puedo hacer yo (Josef)}
% Future work - performance portability analysis

\section*{Acknowledgment}

This research used resources of the Argonne Leadership Computing Facility, a U.S. Department of Energy (DOE) Office of Science user facility at Argonne National Laboratory and is based on research supported by the U.S. DOE Office of Science-Advanced Scientific Computing Research Program, under Contract No. DE-AC02-06CH11357.

\section*{Reproducibility}
    All implementations were run using publicly available code repositories \cite{nbodygit, vortgit}. Scripts are provided for compiling the code, and reproducing the results presented in this study.

\printbibliography

\end{document}